\documentclass{article}
\usepackage{graphicx} 
\usepackage{latexsym} 
\usepackage{amssymb} 
\usepackage{bm}
\usepackage{graphicx}

\begin{document}

\markboth{Sauro Succi} {Analogy between quantum gravity and fluid turbulence}

\title{Analogy between turbulence and quantum gravity: beyond Kolmogorov's 1941 theory}

\author{
Sauro Succi$^* {}^\S$ \\
Istituto per le Applicazioni del Calcolo C.N.R.,\\
 Via dei Taurini, 19 00185, Rome
(Italy),\\
and \\
Freiburg Institute for Advanced Studies,\\
Albertstrasse, 19, D-79104, Freiburg, Germany}

\maketitle


\begin{abstract}
Simple arguments based on the general properties of quantum fluctuations
have been recently shown to imply that quantum fluctuations of spacetime obey
the same scaling laws of the velocity fluctuations in a homogeneous incompressible
turbulent flow, as described by Kolmogorov 1941 (K41) scaling theory.
Less noted, however, is the fact that this analogy rules out the possibility of 
a fractal quantum spacetime, in contradiction with growing evidence in
quantum gravity research. In this Note, we show that the notion of a fractal quantum spacetime
can be restored by extending the analogy between turbulence and quantum
gravity beyond the realm of K41 theory. In particular, it is shown that compatibility
of a fractal quantum-space time with the recent Horava-Lifshitz scenario for quantum
gravity, implies singular quantum wavefunctions.
Finally, we propose an operational procedure, based on Extended Self-Similarity 
techniques, to inspect the (multi)-scaling properties of quantum gravitational fluctuations. 
\end{abstract}

{\it Keywords: Fractal space-time, fluid turbulence, scaling laws}



\section{Introduction}

Quantum Gravity (QG) and Fluid Turbulence (FT) stand out as two major unsolved
challenges in modern theoretical physics.
This is largely due to the fact that, despite their distinct physical nature, they are
both characterized by strong non-linearities, which behinder the development 
of a fully-fledged theory.
As a result, any formal analogy between these two
fields is potentially of interest, since it may permit to put ideas 
and techniques developed in one field, to the benefit of the other.
For instance, the recent analogies between Navier-Stokes and Einstein equations
shed some hope that concepts and methods from the celebrated AdS/CFT duality \cite{ADS1,ADS2}, can be
used to attack fluid turbulence through the formulation of a weakly-interacting gravitational dual .
In this paper, we shall discuss a potential contribution in the reverse direction, namely
the possibility of using hierarchical models developed in the framework of the
phenomenology of fluid turbulence, to the potential benefit of QG research.
Analogies between fluctuating hydrodynamics and quantum gravity have been noticed 
since long.  Visually, the QG-FT analogy was first brought up by Wheeler's poignant description of 
the so-called quantum foam \cite{WHEEL1}, i.e. metric fluctuations which become comparable
in size with the metric background at the Planck scale.  
Subsequently, using simple arguments of general relativity and quantum theory, Padmanabhan
was able to point out operational limits in measuring the position of a particle to
a better accuracy than the Planck length \cite{PADDY}.

Recently, the analogy has been tightened, by showing  
that the quantum fluctuation of spacetime obey the 
same scaling laws of velocity fluctuations of turbulent flows, as described 
by Kolmogorov's 1941 theory \cite{K41}.
The argument goes as follows \cite{NG1,NG2}. 
Consider a wavepacket traveling from a source S to a mirror point M a distance $l=|M-S|$ apart. 
Classically, the distance $l$ is measured simply as $l=c t_r/2$, $c$ being the speed of 
light and $t_r$ the return time at which the signal transmitted at time $t=0$ 
is received back by the source S.
Due to Heisenberg's principle, the distance $l$ is subject to an 
uncertainity $\delta l$, which cannot be made smaller than the wavepacket width 
at time $t_r$, $w(t_r)$. The spread of a quantum wavepacket propagating in free space
is given by $w^2(t) = w^2(0) + \frac{D_q^2 t^2}{w^2(0)}$, where $D_q=\hbar/m$ is the
quantum diffusivity. This expression has a minimum at $w_{min}=\sqrt {2 \; D_q t_{r}}$, and 
consequently, 
\begin{equation}
(\delta l)^2 \ge \frac{D_ql}{c}=l \lambda_c, 
\end{equation}
where $\lambda_c = \hbar/mc$ is the Compton wavelength and numerical factors 
have been taken to unity for simplicity.
Differently restated, 
\begin{equation}
\label{COMPTON}
\frac{\delta l}{\lambda_c} \ge (\frac{l}{\lambda_c})^{1/2} 
\end{equation}
at super-atomic scales $l>\lambda_c$.
Note that the fluctuation scales non-analytically with $l$, i.e.  their gradient 
$g(l) \sim \frac{\delta l}{l}$ would diverge like 
$l^{-1/2}$ in the limit $l \rightarrow 0$.
Of course, such limit makes no physical sense, because UV cutoffs must be taken
into account, which is where gravity takes the stage. 
The gravitational bound reads simply 
\begin{equation}
\label{SCHWARTZ}
\delta l > \lambda_s 
\end{equation}
where $\lambda_s = Gm/c^2$ is the Schwartzschild length. 
This condition ensures that the source S has not melted down into a 
black hole by the time the signal is back. Note that at these scales the fluctuations are even more 
singular, as they don't even go to zero in the limit $l \rightarrow 0$, a manifestation
of gravitational singularities. 
Neither this limit, however, is the relevant one at Planckian
scales, which live exactly at the geometrical mean between the 
Compton and the Schwartzschild scales, $\lambda_p^2 = \frac{G \hbar}{c^3} = \lambda_s \lambda_c$. 
Here, the relevant scaling is the given by the combination of quantum and 
gravitational constraints, that is
$(\delta l)^ 3 >  \lambda_c \lambda_s l = \lambda_p^2 l$, or, differently restated,
\begin{equation}
\label{NG13}
\frac{\delta l}{\lambda_p} >  (\frac{l}{\lambda_p})^{1/3}
\end{equation}
Thus, at the Planck scale we still find a singular behaviour of $\delta l$, with 
an intermediate exponent ($1/3$) between the Schwartzschild ($0$) and Compton ($1/2$) regimes.
Note that in all cases, the gradient $g(l)$ is singular, with exponent $-1$, $-2/3$ and $-1/2$, in
the Schwartzschild, Planck and Compton regimes, respectively.
Interestingly, $1/3$ is exactly the exponent predicted by 
Kolmogorov's 1941 theory (K41) of homogeneous, incompressible turbulence \cite{NG3,SUCCI}. 
More precisely, K41 predicts that the velocity fluctuations 
in a turbulent flow, scale like $\delta v(l) = v_k (l/l_k)^{1/3}$, where 
$l_k$ is the Kolmogorov dissipative length and $v_k \equiv \delta v(l_k)$. 
The analogy is apparent, upon the identification 
$\delta l \leftrightarrow \delta v$ and $\lambda_p \leftrightarrow l_k$.

For fluid turbulence, the $1/3$ exponent results from a specific assumption
on the physical nature of dissipative processes, namely the scale invariance 
of the dissipation rate (energy dissipated per unit volume and unit time), that is: 
\begin{equation}
\label{K41}
\epsilon (l) \equiv \frac{\delta v^2(l)}{\tau(l)} \sim \frac{\delta v^3(l)}{l} = const.
\end{equation}

In \cite{SUCCI}, it was further noted that  the exponent $1/3$ implies
that the set where energy is dissipated is space-filling, ruling out fractals.
Following upon the exact analogy, a similar statement would also apply to the
metric fluctuations, thereby casting questions on the various theories of fractal quantum
spacetime \cite{QFRAC1,QFRAC2}. 
Therefore, either the simple derivation of the $1/3$ law
in quantum gravity is inaccurate, or the fractal theory of quantum spacetime must be revisited.
In view of the mounting evidence of non-integral and scale-dependent effective dimensions
of spacetime \cite{LOLL}, here we pursue the former alternative.
In the following, we shall show that the notion of a fractal quantum spacetime can 
be reconciled with the turbulence analogy, provided the analogy is extended 
to well-known generalizations of the K41 theory.
Most interestingly, such generalizations are shown to be compatible with recent 
statistical field theory formulations of QG, the so-called Horava-Lifshitz picture \cite{HL},
as well as the Dynamic Triangulation (DT) approach to numerical quantum gravity.
The K41 picture is recovered in the IR limit of a smooth 4-dimensional spacetime.
The QG-FT analogy can be "continued" towards the UV scales, on condition of 
replacing K41 with its  well-know generalizations (Kolmogorov's 1961, K61 for short). 

\subsection{Kolmogorov K41 theory,  cascade models and multifractals}

Before addressing these generalized scenarios, let us briefly revisit the way fractal 
sets (fail to) emerge within the standard K41 theory of turbulence.
To this purpose, let us remind the central notion of energy cascade in fluid turbulence.
This refers to the picture of a turbulent flow as a collection of coherent excitations 
(eddies). Under the effect of non-linear interactions, large eddies break down in smaller
eddies, which in turn further break down in smaller eddies, and so on down the line, until
the Kolmogorov length is reached, below which dissipation takes over, thereby terminating
the energy cascade. To be noted that energy cascades from large to small scales 
virtually unchanged, all dissipation taking place at and below the Kolmogorov scale.
The assumption that the dissipation rate be scale-invariant actually implies that dissipation
is a homogeneous process, filling up the entire space occupied by the turbulent fluid.
A very convenient modeling framework to quantify this idea, if only on phenomenological 
grounds, is provided by the so-called hierarchical cascade models \cite{TURBO}.
Within this model, the energy cascade can be cartooned as follows.
A mother eddy at scale $l_0$ breaks down into $2$ eddies of scale 
$l_1=r l_0$, where $r<1$ is an arbitrary fragmentation factor. 
The daughter eddies, in turn, break down into "niece" eddies of size 
$l_2=rl_1=r^2 l_0$ and so on down the line, till the N-th eddy meets the
Kolmogorov scale, $l_0r^N = l_k$, thereby terminating the cascade.
The velocity fluctuations associated with an eddy of $n$-the generation are assumed to scale
like $\frac{\delta v_n}{\delta v_0} = (l_n/l_0)^h$, where 
{\bf $v_0$ is a typical fluid velocity at scale $l_0$} and 
$h$ is an unknown exponent at this point.
Imposing scale-invariance of the energy flux, i.e. $\delta v^3/l=const$, yields 
$3h=1$, singling out $h=1/3$ as a scaling exponent.
Central to this result is the space-filling character of the cascade, i.e. each 
and every fragmented eddy fragments entirely into further daughter eddies
\begin{figure}
\begin{center}
\includegraphics[scale=0.30]{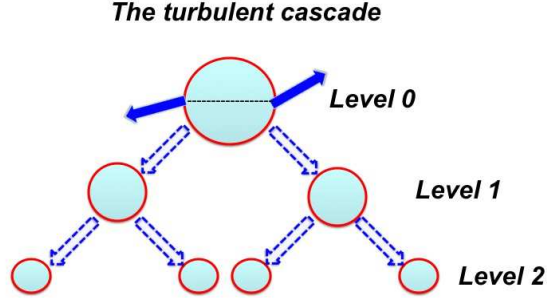}
\end{center}
\caption{The Richardson cascade of turbulent eddies. 
The fluctuating quantity is the velocity fluctuation across the eddy (solid arrows).
For simplicity, this is shown for the mother eddy only.}
\end{figure}
A qualitatively different picture emerges by assuming that only a volumetric
fraction $\beta<1$ of the daughter eddies remains active for further fragmentation.
The dissipation rate at the $n$-th generation is now given by
$\epsilon_n = \beta^n \frac{\delta v_n^3}{l_n}$.
Let us now posit $\beta^n = (\frac{l_n}{l_0})^d$, which {\it defines} 
the parameter $d=D-D_f$ as the defect dimension, i.e. the co-dimension 
of the set where the cascade takes place, $D_f$ being its 
fractal dimension in an embedding space of dimension $D$. 
By its very definition,
$d=\frac{log \beta}{log r}$, from which it is seen that the space-filling scenario, $\beta=1$, 
yields $d=0$, i.e. no fractal set.
By imposing again scale invariance of $\epsilon_n$, we now obtain
$3h-1+d = 0$, i.e. 
\begin{equation}
\label{HD}
h=\frac{(1-d)}{3} 
\end{equation}
This highlights a one-to-one connection between the fractal dimension of the cascade space 
and the scaling exponent of the velocity fluctuations which live on it.
In passing, we note that the most violent fluctuations take place
on sets with large co-dimensions $d$, i.e. small fractal dimension $D_f$.
In fluid turbulence, the most singular structures are typically credited for being
filaments of fractal dimension $D_f=1$, $d=2$ and $h=-1/3$.

It is well known that real-life turbulent flows do {\it not} obey 
the relation (\ref{HD}), i.e. they are not described by a single 
fractal, of whatever dimension.
Experimentally, this is revealed by inspecting the so-called 
structure functions of order $p$:
\begin{equation}
\label{SP}
S_p(l) = < \delta v^p(l) > = (\delta v_0)^p (\frac{l}{l_0})^{\zeta_p}
\end{equation}
where brackets stand for ensemble averaging over turbulent realizations
and $\zeta_p$ are the corresponding scaling exponents of order $p$.
By construction, high $p$'s probe rare and highly energetic events (bursts).
{\bf Should turbulence be described by a single fractal, as per eq. (\ref{HD}), one
would observe a linear dependence of the scaling exponents} on $p$, namely
$\zeta_p = hp$. This is contradicted by experimental evidence, which shows
instead a sub-linear dependence of the form  
\begin{equation}
\label{ZP}
\zeta_p = hp + \eta_p
\end{equation}
where $\eta_p<0$ describe the so-called "intermittency" corrections.
Intermittency refers to the fact that, as suggested by visual 
experience, turbulence is all but a homogeneous process. 
Quite on the contrary, dissipation typically comes through spotty bursts and gusts, which 
stand in stark contrast with the notion of a scale-independent, space-filling, homogeneous 
dissipation rate. This observation motivated the development of Kolmogorov's 
1961 theory \cite{K61}.
By using the definitions (\ref{K41}) and (\ref{SP}), we readily obtain
$S_p(l) = < \epsilon^{p/3} l^{p/3} > \sim l^{p/3} l^{\tau_{p/3}}$, where $\tau_p$
are the scaling exponents of the dissipation field.   
Comparison with (\ref{ZP}), shows that $\eta_p \equiv \tau_{p/3}$, i.e. deviations
of the scaling exponents $\zeta_p$ from linear behaviour can be ascribed to the 
fluctuations of the dissipation rate.
A powerful notion to account for intermittency, if not explain 
it, is the concept of multifractal, whereby turbulent dissipation takes place on a 
sequence of fractals (multifractal), each with its own fractal 
co-dimension $d(h)$, varying within a continuous range $h_1 \le h \le h_2$.
Although multifractals do not provide an explanation for intermittency, they set
nonetheless a powerful mathematical stage for a geometrical theory of turbulence \cite{MUFRA}. 
Here, we conclude this brief excursus by noting that, at level of cascade modeling, the
notion of multifractal is readily incorporated by promoting the volumetric fraction
$\beta$ from a mere parameter to a random distribution (random beta model).

\subsection{Quantum gravity inverse cascade}

Back to quantum gravity, the question is whether the QG-FT analogy survives the extensions
of Kolmogorov's theory, and, if so, how does one accomodate the notions of intermittency,
multifractals and associated cascade models.
In the sequel, we shall show that the analogy can indeed be taken to this
extended territory, provided quantum spacetime fluctuations are treated like
a fractional brownian motion, in line with recent theories of 
anisotropic spacetime and quantum gravity at a critical Lifshitz point \cite{HL}.   
Borrowing the cascade language for quantum gravity, one postulates an {\it inverse}
cascade (UV to IR), whereby two small mother eddies of size $l_0$ coalesce into a single 
daughter eddy of size $l_1 = l_0/r$ ($r<1$) and so on, {\it up} the line.
The formalism applies all the same, in reverse, with the obvious duality
$r \rightarrow 1/r$ and $\beta \rightarrow 1/\beta$, which leaves $d$ unchanged.
Thus, if metric fluctuations live in a fractal spacetime of dimension $D_f=D-d$, their 
scaling exponent is $h=(1-d)/3$, violating the simple scaling relation eq. (\ref{NG13}). 
Since many evidences are now accumulating for a fractal spacetime, with spectral dimension 
ranging from about $2$ at high-energy, up to the full $4$ dimensions at low 
energy \cite{LOLL}, it makes sense to revisit the main procedure leading to the $1/3$ exponent.
To this purpose, let us assume that the signal propagation
obeys a generalized Brownian motion of the form $dl = c_z dt^{1/z}$, where $z$ is the 
dynamical exponent at a critical Lifshitz point.  
The case $z=1$ recovers standard advection (smooth manifold), and 
$z=2$ to standard diffusion (non-differentiable manifold).
The recent Horava-Lifshitz (HL) proposal of anisotropic spacetime fluctuations
(Lifshitz critical point), predicts a spacetime spectral dimension 
$D_s=1+D/z$, in $D$ spatial dimensions.
More precisely, the HL theory predicts $z=3$, i.e. $D_s=2$ in the UV range, up 
to $z=1$, i.e. $D_s=4$ at large scales.  
In our language, $D/z \equiv D-d$, i.e. $d=D(1-1/z)$, so that $z=3$ and $z=1$
yield $d=2$ and $d=0$ respectively in $D=3$, corresponding to spectral dimension
$D_s=2$ and $D_s=4$ respectively, in close match with  
numerical quantum gravity simulations based on causal dynamic 
triangulations (CDT) of spacetime \cite{LOLL3,VISSER}.  
To be noted that the corresponding exponents run from $h=1/3$ (IR) to
$h=(1-2)/3=-1/3$, pointing to a very singular UV behavior of the metric fluctuations.
These fluctuations are wilder than turbulent velocity fluctuations, but less
violent than the fluctuations of the dissipation field $\epsilon(l) \sim l^{-2/3}$.
In this sense, the QG-FT analogy still holds, although more in relation to
flow dissipation than the flow velocity itself.  
Assuming that the minimum spread of the wavepacket still obeys the Schroedinger scaling
$\delta^2_{min} \sim Dt$, the Schwartzschild constraint $\delta l > \lambda_S$, as combined
with a HL return time $t_r \sim l^z$, would yield 
$(\delta l)^3 \sim l^z$, that is $\delta l \sim l^{z/3}$, i.e. $h=z/3$.
With $z=1$, this returns the familiar exponent $h=1/3$, while $z=3$ yields $h=1$, in stark contrast 
with the previous finding $h=-1/3$.
Clearly, some assumptions need to be revisited.
In particular, in a fractal spacetime, there is no reason to believe that
the spatial spreading of the wavepacket should obey the same diffusion-like
relation as in smooth spacetime.
Assuming that in the HL spacetime the minimum wavepacket spread obeys a scaling
relation of the form $\delta^2_{min} \sim t_r^{\alpha(z)}$, combination with $t_r \sim l^z$,
would give $h=\alpha(z) z/3$. Clearly, in the limit $z \rightarrow 1$ 
we require $\alpha \rightarrow 1$, so as to recover the standard K41 scenario.
Compatibility with the relation $h=(1-d)/3$ and the identity $d=D(1-1/z)$, yields
a spreading exponent 
\begin{equation}
\alpha(z) = \frac{1}{z} - D (\frac{1}{z}-\frac{1}{z^2}) 
\end{equation}
This relation is reported in Figure 2 as a function of $z$ for the case $D=3$.
First, we note that indeed $\alpha$ goes to the unit value for $z=1$, as it should.
Second, we observe that the spreading exponent becomes negative for 
$z>z^{\star}(D)=\frac{1}{1-1/D}$, namely $z^{\star}=3/2$ for $D=3$.
Third, the minimum exponent is attained at $z=z_{min}(D)=\frac{2}{1-1/D}=2z^{\star}$,
its value being $\alpha_{min}=-\frac{(D-1)^2}{4D}$.
Amazingly, for $D=3$, the minimum value, $\alpha_{min}=-1/3$ is attained precisely
at the HL value $z=3$.
Finally, and most importantly, we observe that a negative spreading exponent implies
the development of a finite-time singularity in the quantum wavefunction carrying the
propagating signal.
Singular wavefunctions are known since long in quantum mechanics \cite{BERRY}, however
to the best of this author knowledge, they have not been discussed before in the framework
of quantum gravity.
\begin{figure}
\begin{center}
\includegraphics[scale=0.4,angle=0]{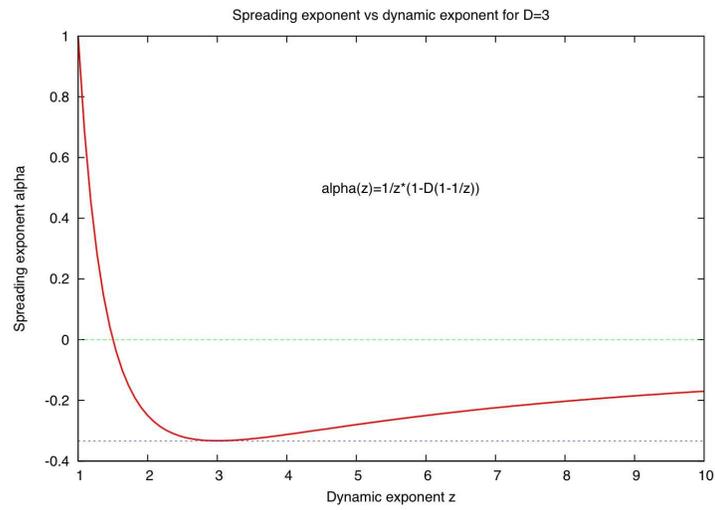}
\end{center}
\caption{The spreading exponent of the quantum wavefunction, $\alpha$, as a function of the
dynamic exponent $z$ for the case of three spatial dimensions $D=3$.
To be noted that, for $z>3/2$, the spreading exponent becomes 
negative, pointing to the development of a finite-time singularity in 
the quantum wavefunction carrying the propagating signal. 
}
\end{figure}

\subsection{Extended Self Similarity for quantum gravity}

The existence of a continuous range of spectral exponents is conducive to
the idea of QG multifractals. 
In order to detect them, it appears natural to define 
QG structure functions and associated scaling exponents:
\begin{equation}
\label{GP}
G_p(l) = < (\frac{\delta l}{\lambda_p} )^p> \sim (\frac{l}{\lambda_p})^{\gamma_p}
\end{equation}  
These could be measured in QG simulations, possibly using the techniques
of extended self-similar similarity, to be detailed shortly \cite{ESS1,ESS2}.
In the sequel, we provide an operational procedure to accomplish this task.
Let us write the volume change from scale $l$ to $l+\delta l$ as
\begin{equation}
\label{VOLFLU}
\delta V(l) \equiv V(l+\delta l) - V(l) = 
v_1 l^2 \delta l + v_2 l (\delta l)^2 + v_3 (\delta l)^3
\end{equation}
where $v_k$, $k=1,3$ are $O(1)$ geometry-dependent constants.
{\bf
For a cube $v_1=v_2=3$ and $v_3=1$ and for a sphere the same sequence is
pre-factored by $4 \pi/3$. 
}
For a generic tetrahedron in a CDT realization
of spacetime, $v_k$ should be taken as random numbers, whose statistics
contains all the non-perturbative physics beyond simple scaling, if any.

For a generic scaling law of the form
$\frac{\delta l}{\lambda_p} \sim (\frac{l}{\lambda_p})^h$,
one readily computes
$y \equiv \frac{\delta V(l)}{V_p} = 
v_1 x^{2+h} + v_2 x^{1+2h} + v_3 x^{3h}$
where $h<1$ is the scaling exponent, $V_p=\lambda_p^3$ is the
Planck volume, and we have set $x \equiv l/\lambda_p$.
If the moments of $v_k$ remain finite at any order, then $\gamma_p = (2+h)p$,
otherwise, multifractal behaviour is observed.
For $x>>1$, the first term on the rhs leads the series, hence the
scaling exponent of volume fluctuations is simply $2+h$.
One could obtain $\gamma_p$ by log-plotting $G_p$ versus
$x$, using the data from a CDT simulation.
In actual practice, however, the scaling relation $y \sim x^{2+h}$ usually holds only
in a restricted interval $1<<x<<\Lambda \equiv L/\lambda_p$, usually too
small to allow statistical accuracy (present day CDT simulations
feature $\Lambda \sim 10^2$). 
While waiting for computer advances, one must turn to alternative tools.
One which turned out to be pretty useful to unravel the scaling properties
of turbulent flows is Extended Self Similarity (ESS).
ESS maintains a generalized scaling law of the form
\begin{equation}
\label{ESS}
y \sim \psi(x)^{2+h}
\end{equation}
where $\psi(x)$ is a universal function of $x$, which reduces to $x$ 
only in the limit $x>>1$ (one can think of it as of a generalized space
coordinate $x'=\psi(x)$, probing a larger set of scales).
The ESS scaling (\ref{ESS}) then holds on a broader range than the
native scaling $y \sim x^{2+h}$.
Unfortunately, $\psi(x)$ is generally not known a-priori, and consequently
ESS cannot be used to deduce the exponent by log-plotting $y$ vs 
$\psi(x)$ from numerical data.
However, {\it relative} scaling functions can be used to bypass the problem.
It is indeed clear that under ESS conditions, the following relative 
scaling between two generic structure functions of order $p$ and $q$, holds
\begin{equation}
G_p \sim (G_q)^{\gamma_p/\gamma_q}
\end{equation}
where $\gamma_p/\gamma_q$ reduces to $p/q$ for the case of simple scaling
(no multifractal).
In case the scaling exponent is analytically known for some reference
value $q^*$, relative scaling of $G_p$ versus $G_{q^*}$, uniquely
delivers $\gamma_p$. In fluid turbulence $q^*=3$ and $\gamma_{3}=1$.
We are not aware of a corresponding QG analogue.
The advantages of ESS analysis are clear: i) no knowledge of 
$\psi(x)$ is required, ii) the extraction of
$\gamma_p$ from numerical data can rely on a broader range of values.
These have proven especially valuable in low-resolution experiments in fluid turbulence, where 
standard scaling would not hold on a sufficiently broad range of wavenumbers.
The same advantage is expected apply to the statistical data from CDT simulations.
The procedure described above is unambiguous, hence fully operational 
once a set of CDT data is available

\section{Conclusions}

Based on the quantitative analogy between the scaling exponent
of quantum spacetime fluctuations and fluid turbulence within Kolmogorov
K41 theory, we conclude that the common exponent $h=1/3$ rules
out the possibility of a fractal quantum spacetime.
In this Note, we have shown that such a possibility can be restored by moving to the generalized
Kolmogorov 1961 theory and drawing a parallel with a critical Horava-Lifshitz QG 
scenario, whereby  space an time would fluctuate with different exponents.
Within such generalized scenario, QG fluctuations look more akin to the
fluctuations of the dissipation rate rather than to velocity fluctuations and imply
the developmemnt of finite-time singular quantum wavefunctions.
Finally, an operational procedure to measure the (multi)-scaling properties of quantum gravitational
fluctuations, based on Extended Self Similarity techniques, has been suggested.

\section{Acknowledgements}
Illuminating discussions with B.L. Hu, Y. Oz and T. Padmanabhan, during the European Science Foundation
Exploratory Workshop "Gravity as Thermodynamics" (EW10082), are kindly acknowledged.
I am also grateful to T. Padmanabhan for valuable comments and remarks on a preliminary version
of this manuscript.



\begin{thebibliography}{99}

\bibitem{ADS1}  J. Maldacena, Adv. Theor. Math. Phys. 2, 231 (1998),

\bibitem{ADS2} O. Aharony, S. S. Gubser, J. M. Maldacena, H. Ooguri and Y. Oz,
 Phys. Rept. 323 (2000) 183, [arXiv:hep-th/9905111].

\bibitem{WHEEL1} J. Wheeler, Ann. Phys. NY, 2, 604, (1957)

\bibitem{PADDY} T. Padmanabhan, Class. Quantum Grav., 4, L107, (1983)

\bibitem{K41} A.N. Kolmogorov, Dokl. Akad. Nauk USSR, 30, 9 (1941)

\bibitem{NG1} Y. J. Ng, H. van Dam, Mod. Phys. Lett. A., 9, 225 (1994)

\bibitem{NG2} Y. J. Ng, Phys. Rev. Lett., 86, 2946 (2001)

\bibitem{NG3} V. Jejjala et al, Classical and Quantum Gravity, 25, 225012 (2008)

\bibitem{SUCCI} S. Succi, Int. J. Mod. Phys. C, (2010)

\bibitem{QFRAC1} L. Nottale, Int. J. Mod. Phys. Am 19, 5047 (1989)

\bibitem{QFRAC2} D. Benedetti, Phys. Rev. Lett. 102, 111303 (2009)

\bibitem{LOLL} J. Ambjorn, J. Jurkiewicz, R. Loll,  Phys. Rev. Lett., 93, 131301 (2004);

\bibitem{HL} P. Horava, Phys. Rev. Lett. 102, 161301, (2009)

\bibitem{TURBO} U. Frisch, Turbulence, Cambridge U.P., (1996)

\bibitem{K61} A.N. Kolmogorov, J. Fluid Mech., 83, 13 (1962)

\bibitem{MUFRA} R. Benzi et al, J. Phys. A, 17, 3521, (1984) 

\bibitem{LOLL3} J. Ambjorn, A. Goerlich, J. Jurkiewicz, R. Loll, Phys. Rev. Lett., 100, 091304 (2008);

\bibitem{VISSER} T.P. Sotiriou, M. Visser and S. Weinfurtner, Phys. Rev. Lett., 107, 131303, (2011)

\bibitem{BERRY} M. Berry, Rep. Prog. Phys., 35, 315 (1972).

\bibitem{ESS1} R. Benzi, S. Ciliberto, R. Tripiccione, F. Baudet, F. Massaioli
and S. Succi, Phys. Rev. E, 48, R29, (1993).

\bibitem{ESS2} M. Briscolini, R. Benzi, P. Santangelo and S. Succi, 
Phys. Rev. E, 50, R1745, (1994) 

\end{thebibliography}
\end{document}